# Epitaxial graphene on SiC: 2D sheets, selective growth and nanoribbons

Claire Berger[1,2], Dogukan Deniz[1], Jamey Gigliotti[1], James Palmer[1], John Hankinson[1], Yiran Hu[1], Jean-Philippe Turmaud[1], Renaud Puybaret[3], Abdallah Ougazzaden[3], Anton Sidorov[1], Zhigang Jiang[1], Walt A. de Heer[1]

[1.] *School of Physics, Georgia Institute of Technology, Atlanta, GA 30332, USA*
[2.] *Institut Néel, CNRS- Université Grenoble Alpes, 38042 Grenoble, France*
[3] *Georgia Institute of Technology - CNRS UMI 2958, 57070 Metz, France*

**Abstract**
Epitaxial graphene grown on SiC by the confinement controlled sublimation method is reviewed, with an emphasis on multilayer and monolayer epitaxial graphene on the carbon face of 4H-SiC and on directed and selectively grown structures under growth-arresting or growth-enhancing masks. Recent developments in the growth of templated graphene nanostructures are also presented, as exemplified by tens of micron long very well confined and isolated 20-40nm wide graphene ribbons. Scheme for large scale integration of ribbon arrays with Si wafer is also presented.

**Introduction**

Fifteen years of research on epitaxial graphene on SiC (hereafter called epigraphene) have largely demonstrated its potential not only as the best graphene nanoelectronics platform but also the best platform for a large variety of basic science studies of graphene.[1] From its inception in 2001, it was realized that graphene for nanoelectronics needs an atomically well defined substrate. This requirement is satisfied by growing the material on single crystal substrates. Previous surface science studies of single layer graphite grown on hexagonal SiC substrate had demonstrated that it has characteristics of isolated graphene. In general, large scale integration calls for commercially available wafer scale substrates and best graphene science is demonstrated on flat, unstrained and un-rippled surfaces; all this makes SiC a substrate of choice for graphene. SiC is widely available at affordable price, as is clear from its widespread use as a substrate for LED lighting.

Epigraphene production methods have been developed allowing not only 2D sheets[2-11], but also intricately patterned structures to be grown on the substrate[12, 13]. The ability to grow graphene nanostructures cannot be underestimated: if graphene is to be used in nanoelectronics, atomically defined graphene structures need to be patterned at precise locations. The methods presented here[12] [14-16] (masking and more especially template growth on sidewall of SiC trenches) circumvent the detrimental effect of traditional plasma etching methods, and resulting nanoribbons show exceptional room temperature ballistic transport properties[17].

Furthermore, to achieve the best performance graphene needs to be annealed and encapsulated in order to mitigate the effects of adsorbates from exposure to the ambient atmosphere. An advantage of epigraphene is that it is resistant to high temperature and harsh radiation conditions. This review will focus on two main aspects of epigraphene material. First we will outline the quality of epigraphene 2D sheets when grown by the confinement controlled sublimation method; then we will present alternative methods to pattern graphene at the nanoscale with an emphasis on structured growth, and show paths toward large scale integration.

## 1. Near equilibrium Confinement Controlled Sublimation growth

The growth of graphene by SiC decomposition is non-conventional in the sense that it doesn't require an external source of carbon. Rather, at high temperature and in vacuum, the SiC surfaces decompose by silicon sublimation, resulting in a carbon rich surface that forms a graphene layer on the SiC[18]. The key to the growth of high crystalline quality graphene layer is to control the rate of silicon escape $\Gamma_{Si}$ from the surface, which is a balance between the rate of sublimation from and absorption onto the surface (Figure 1a)[3]. $\Gamma_{Si}$ is controlled by the temperature and the background pressure of silicon vapor. When the SiC surface is in equilibrium with the Si vapor ($\Gamma_{Si}=0$), the formation of graphene is arrested.

The Confinement Controlled Sublimation (CCS) growth method relies on a near equilibrium growth condition by controlling $\Gamma_{Si}$. Details can be found in Ref. 3. A SiC crystal is placed in a graphite enclosure (the crucible). During annealing the evolving Si vapor escapes through a small calibrated hole (see Figure 1b). A background of neutral gas pressure may be supplied that further slows down silicon diffusion. The graphite crucible is placed in a vacuum chamber. The system is pumped down to a base pressure of $10^{-7}$ mbar with a turbo-molecular pump, then uniformly heated using RF coils that heat the crucible, as shown figure 1c, to temperatures that can reach up to 2100°C. The system is designed to have no heating elements inside the vacuum chamber, so that the SiC surface is exposed only to its own vapor. In our most recent furnace design[19], the induction susceptor is the graphite crucible itself. It is supported directly by the quartz tube which temperature remains well below its melting point even for the highest crucible temperature (2100ºC for more than 20 minutes), owing to the small actual contact area between the tube and the crucible. The compact design allows for fast pumping speed. The temperature is measured externally with a two-wavelength pyrometer and the assembly has very little thermal inertia, so that fast heating and cooling rate can be achieved up 150°C/s in the whole temperature range. The temperature profile is completely computer automated with temperature ramps and plateaus fully configurable by the user. The RF-power is adjusted using a proportional-integral-derivative controller that manages the feed-back loop between the measured and set temperatures. We have estimated[19], that gas pressure equilibration is reached in the crucible within 5µs, which is many orders of magnitude shorter than a typical growth time. This confirms that growth occurs very close to equilibrium. Further, since solid/ vapor pressure is an intensive property that does not scale with surface area when in equilibrium, the SiC chip area inside he crucible does not affect the growth rate.

To grow graphene on hexagonal 4H or 6H-SiC wafers a typical temperature cycle includes a degasing stage at 750°C-800°C for a few minutes, followed by a SiC surface structuring stage around 1100-1250ºC before the Si sublimation at high temperature (1400-1650°C ). Optimum temperature and growth times are empirically determined, and stable recipes have been established for each type of structure, including graphene buffer layer, mono and bilayer on the Si-face, from monolayer to thick (more that 50 layers) on the C-face (see Figure 1d), and for confined growth on other than on-axis facets (for instance SiC sidewalls, see below). The exact temperature profile depends on the background pressure (vacuum or 1 atm Ar), the geometry (explicitly the hole size) and condition of the graphite crucible. We have noted a slow evolution with time of the growth parameters, which is due to a change in the adsorption of Si on graphite crucible walls after many growth cycles. The original conditions can be restored by baking the graphite crucible at high temperature.

### 2.1. Multilayer C- face

The near-equilibrium growth condition of the CCS method is favorable to grow very high quality epigraphene films. On the 4H/6H-SiC carbon face, typically 5-10 layers films are easily grown[20] and more than 40 layers can be produced[21], while single layer films are much harder to achieve[22]. The number of layers is determined by both temperature (in the range 1450–1525°C) and time (1 minute to several 1 hour cycles for very thick films).

Figure 2 presents atomic force microscopy (AFM) and scanning tunneling microscopy (STM) images of few layer graphene on the C-face. At high resolution the graphene honeycomb lattice is revealed (Figure 2a), and large-area STM scans[23] show that the top layer is extremely flat (Figure 2b) on terraces of tens of μm size (Figure 2c) in well formed C-face graphene. Note the vertical scale of less than 0.1 nm in the 400 by 400 nm STM scan of Figure 2b and the very low roughness (rms< 50pm) on the trace in Figure2d that extends over regions of different moiré[24]. Extremely low roughness values are confirmed by x-ray diffraction (rms<5pm)[25]. No grain boundaries have been seen in large-area STM scans (Figure 2d), indicating the top most graphene layer is continuous over the whole sample, which makes C-face epigraphene among the largest synthetic 2D crystal. The moiré patterns seen in Figure 2d provide clear evidence that adjacently stacked graphene layers are rotated creating a superstructure. Variations in the relative rotation of the layers change the moiré from place to place, with no change in the topographic height of the top layer (Figure 2d). On a larger scale, smooth graphene pleats are observed, as seen as bright lines in Figure 2c (see also SEM image of Figure 2e). The pleats (also referred to as puckers, ridges, wrinkles, folds, ripples, etc in the literature) result from the larger thermal contraction of SiC compared to graphene when the films are cooled down from the growth temperature (around 1500-1600°C) to room temperature.

The rotational layer stacking in C-face graphene is new and explicitly is not turbostratic (random stacking of small grains). In CCS grown C-face, the stacking as determined by x-ray diffraction alternates $0\pm\delta\theta$ degree/ $30\pm \delta\theta$ degree oriented layers, with a variation $\delta\theta$ within typically a few degrees.[25] Graphite (AB or rhombohedral) stacking in this case corresponds to stacking faults that have an occurrence of less than 15-19%. [25]

The properties of well grown CCS C-face multilayered epitaxial graphene (MEG) are exceptional and have been amply discussed in the literature (see for example review 1). It was soon realized that MEG is a model system to study graphene properties. This stems from a combination of factors: (i) Excellent structural quality, with very few defects and continuous sheets as shown by the absence of a D peak in Raman spectroscopy and by STM scans. (ii) Flatness of the layers, which excludes strain related gauge field effects and broadening effects in k-resolved surface spectroscopy measurements, such as Angle Resolved Photoemission Spectroscopy (ARPES). (iii) Extremely small interaction with the substrate and the environment, more specifically for the layers in the middle of the stack that are quasi-neutral (charge density n~$5 \times 10^9$ cm$^{-2}$, that is at most 8 meV away from the Dirac point). (iv) Rotational stacking resulting in each layer in the stack having the electronic structure of monolayer graphene, explicitly not graphite.

Most prominently, the quasineutral layers show record high mobilities of $10^6$cm$^2$/Vs at room temperature, as determined by infra-red magnetospectroscopy with no sign of dependence with temperature[26]. Transport measurements that include contribution from the negatively doped layer at the interface consistently show high (but reduced) mobilities of the order of 15,000 to 25,000 cm$^2$/Vs,[20, 27] and characteristics of graphene such as a Berry's phase of $\pi$ and weak anti-localization.[1] Spin transport is highly efficient in MEG with record spin diffusion lengths up to 285µm[27]. These layers show textbook graphene electronic structure that allow spectroscopic studies (STS, ARPES, magnetospectroscopy) down to instrumental resolution.[24, 26, 28] Most notably, the fine structure of the Landau levels and their real space mapping was observed for the first time on a 2 dimensional gas system by STS measurement of the top C-face epigraphene layer. MEG is also very much sought for optical measurements due to the large area and homogeneity of the graphene layers residing on a transparent substrate. Moreover, since MEG is made of effectively decoupled graphene layers, the multiplicity enhances the signal without compromising the graphene characteristics, contrary to thin graphite. As an example of compelling result obtained from ultrafast optical spectroscopy studies in MEG the electron dynamic probed reveals the details of electronic relaxation in graphene, as an interplay between efficient carrier-carrier and carrier-optical-phonon scattering, and interlayer energy transfer effects.[29]

*2.2. Monolayer C-face*

Monolayer graphene is more difficult to produce due to the rapid growth rates on the C-face and the multiple nucleation sites. Usually isolated graphene patches up to tens of µm in size are produced by the CCS method [13, 22, 30] at a graphitization temperature around 1500°C. The graphene patches extend over multiple steps and show the characteristic pleat structure of C-face graphene (see Figure 3a). With optimized growth conditions (typically around 1500°C for 10 min) larger patches and extended monolayer sheets can be produced as seen in AFM image of Figure 3a. An alternative method to CCS is to place a piece of graphite on the SiC chip, which provides some Si vapor confinement. This way narrow tapered stripes of monolayer graphene were produced[5].

The monolayer nature of the CCS produced graphene is assessed with combined AFM imaging and Raman spectroscopy (narrow 2D peak), as shown in Figure 3b. Raman spectroscopy is widely used to characterize graphene. The symmetrical and narrow graphene 2D peak observed in epigraphene, provides an unambiguous signature of graphene monolayers. The D and G peaks are in the same energy range as the second order SiC Raman peaks. A simple subtraction of the SiC contribution to the total spectra is generally performed but for monolayer graphene the similar intensity of the SiC and graphene peaks results in poorly defined graphene spectra. The Non-negative Matrix

Factorization (NMF) decomposition technique [31] allows the pure graphene and SiC spectra to be segregated from the raw data. This is done by decomposing a set of spectra, taken at different focal points of the exciting Raman laser light at, and just below the surface, into a final number of spectral components. Details of the method are provided in [31]. A notable advantage of the technique is that the decomposition does not require prior knowledge of the components spectral profile, or even of the number of components needed (two in the case of graphene on SiC). The method applies to separate the contributions of stacked layers (such as graphene on SiC) or the contribution of patched films in inhomogeneous 2D maps. An example is given in Figure 3b, where the Raman spectrum of single layer graphene is extracted from the raw data. Significantly, the D peak is very small, attesting the good graphene quality.

The epitaxial graphene monolayer is further unambiguously characterized from the observation of the quantum Hall effect (QHE),[13, 22] as shown in Figure 3c –d for two samples of high mobility µ = 39,800 cm$^2$/Vs (20,000 cm$^2$/Vs) and charge density n = 0.19×10$^{12}$cm$^{-2}$ (0.87 ×10$^{12}$cm$^{-2}$), respectively. Note that unlike the graphene layers in the middle of the stack in MEG that are quasi neutral, the graphene layer at the interface with SiC is naturally n-doped with a few 10$^{12}$cm$^{-2}$. Positive counter-doping, either by exposure to the ambient or with a top electrostatic gate is necessary to reduce the charge density close to the Dirac point.[22] The samples in Figure 3 show two attractive characteristics of graphene for resistance metrology standard based on the quantum Hall effect, namely a robust Hall resistance quantization starting at low magnetic field and still observed at high temperature. The quantized $\rho_{xy}$=12.8kΩ=(2e$^2$/h)$^{-1}$ Hall plateau in Figure 3c has an extension of more than 4 Tesla with an onset below 3 Tesla, that is within the range of commercial room temperature electromagnets. A very well defined plateau is observed at 100K that is above the widely available liquid N$_2$ temperature (Figure 3d).[32] From a metrology perspective, note that the most accurate QHE plateau quantization that rivals the best GaAs 2DEGs QHE standards was measured in epigraphene[33, 34].

Having established a reliable production method of high mobility monolayer graphene on the C-face, the samples can be used for field effect transistors (FETs).[35, 36] For this the samples are patterned in a dual source/gate and common drain structure (Figure 3e inset) that is designed for ultra-high frequency measurements. By having contacts and T-shape gates optimized to minimize access resistances and parasitic capacitances, C-face graphene FETs show record maximum oscillation frequency $f_{max}$=70GHz.[35] The frequency $f_{max}$ quantifies the practical upper bound for useful circuit operation. It was the first time that a power amplification was achieved at a frequency comparable to the current amplification (characterized by $f_T$=110GHz).

### 2.3. Monolayer Si-face

Epigraphene on the Si-face is the most studied epigraphene form in the literature. Growth on the Silicon face proceeds from the SiC steps. This adds an inherent difficulty to produce extended monolayer sheets, because bilayers (or multiple layers) tend to form at step edges when full terrace graphene coverage is reached. As a matter of fact, the step edges act as electronic scattering centers, which has been related to a reduction of the carrier concentration on the sidewall of step, to a n/p junction or the presence of a bandgap at the step edge, or to the presence of bilayers/ multilayers (see Ref. [1] and refs therein). The later have a detrimental effect to the homogeneity of the quantum Hall effect. Better control of epigraphene growth on the Si-face has been obtained by providing carefully balanced methane and $H_2$. [8]

The fact that growth starts at a sidewall, can be tuned to an advantage by arresting the growth on the sidewall of steps (either natural steps or etched steps in SiC - see below in section 3.2). Another consequence of graphene growth initiated at steps is that the quality of the SiC surface prior to growth is essential. Rough surfaces result in inhomogeneous and defective graphene layers, which can be rationalized by the presence of multiple graphene nucleation centers for surface decomposition. Surface flattening (removal of polishing scratches) can be obtained by SiC etching at high temperature (typically 1600ºC) in a hydrogen environment (we use 3% $H_2$ in Argon). Commercial SiC wafers are also available with surfaces treated by a chemical mechanical process, which provide good starting surfaces. For natural steps, another consideration is the width of the terraces that depends on the step bunching for a given miscut. We have used pre-growth annealing in various conditions to help organize the step-terrace structure for a particular desired surface state. These include annealing at moderate temperature (1100°C-1200°C) in vacuum, high temperature in argon, face-to-face SiC surface re-structuring and step pinning under an evaporated refractory cap. In some cases it can be beneficial to grow first the buffer layer (that is the semiconducting graphene layer bound to SiC on the Si-face) to help reduce the step displacement. An example of pre-growth SiC surface structuring is given in Figure 4a, featuring large atomically flat terraces with straight steps. The terraces are 20-40μm in width and extend over hundreds of μm in length (see also Figure 7e).

The capping technique consists in evaporating an amorphous carbon grid onto bare SiC. Upon thermal treatment the SiC steps are pinned under the amorphous carbon and steps displacement is confined to within each amorphous carbon enclosure.[37] This results in step bunching at one end of the enclosure and step alignment along the amorphous carbon corral, as demonstrated in Figure 4b-c. This technique demonstrates a significant improvement in SiC surface structuring by providing not only flat terraces of large size (Figure 4c -bottom) but most importantly at a predefined location. Note also that amorphous C is easily removable by plasma etching after annealing.

## 2. Selective graphene growth

### 3.1. Masking techniques

Any graphene electronic application requires that graphene patterning. The principle of structured growth stems from the realization that patterning a 2D graphene sheet by the usual oxygen plasma removal of carbon atoms is very destructive and leaves roughed edges and uncontrolled orientation and termination. It is therefore advantageous to grow graphene directly at preset locations. Figure 5a-b presents two examples where masking methods were used to grow graphene into shape with no need for further etching. Graphene growth can be either stopped under an AlN mask [14], slowed down under a Si-poor $Si_{3-x}N_4$ mask or promoted under a rich $Si_{3+x}N_4$ mask[15]. In Figure 5a, an 80nm thick AlN film was evaporated then patterned revealing bare SiC in the shape of a multiprobe Hall bar. After annealing (20 minutes at 1420°C) multilayer graphene was grown on the 4H-SiC C-face only on the exposed SiC, while no graphene grows under AlN, as shown by the Raman 2D map of Figure 5a (bottom). In the case of the SiN mask, contrary to AlN, the mask vanishes during the heat treatment, so that there is no need for further mask removal; a differential in graphene growth is revealed where the mask was present. The four panels of Figure 5b show how the Si –rich $Si_{3+x}N_4$ patterned mask (a) was transferred into enhanced selective graphene growth as shown in the optical image (b), the Raman 2D peak (c) and 2D peak/G peak maps (d) with sub-micron resolution. Conversely graphene was used as a mask to prevent GaN growth in Figure 5c. Here holes 75 nm diameters are patterned in MEG (bright dots in the inset) and GaN crystals (triangles in the main panel) grow only on SiC in the holes. This structure is interesting to integrate GaN LED with graphene.

*3.2. Sidewall facets*

The concept of selective growth was applied to grow predefined nanoscale structures at desired locations without masking.[12] The technique relies on the fact that graphene grows faster on facets with orientation other that the (0001) basal plane of the Si-face, as was readily observed by the growth of multilayer graphene at step edges (see above). Facets can be the natural steps arising from the miscut angle relative to the (0001) plane; step bunching by annealing in controlled condition produces arrays of steps and terraces with rather uniform width and height. More interestingly, trenches of various shapes can be etched in SiC, [13] the sidewalls of which recrystallize into the crystallographic equilibrium facets of SiC (like (2-207)-facets) upon annealing. Trenches can be etched for instance along the 4H-SiC-(-1-120) and 4H-SiC-(1-100) directions in SiC providing graphene ribbons along the zigzag or armchair orientation, respectively owing to the epitaxial orientation of graphene on SiC (see figure 6a).

As for natural steps, graphene grows preferentially on these facets; under controlled growth conditions at temperature around 1600°C with the CCS method, graphene growth can be arrested on the sidewalls producing monolayers on them. The ribbon width is then determined by the step height and the facet angle from the basal plane (about 27 degrees). Because the steps in SiC are etched and annealed prior to graphene growth, the facets onto which graphene grows are smooth and atomically defined. This is best demonstrated in cross-sectional transmission electron microscopy (TEM). Figure 6b shows that a single graphene layer drapes over the (here armchair) step and merges into the buffer layer that is tightly bound on the (0001) surface. Significantly, the graphene layer on the main facet is at a significantly larger distance from SiC than the buffer layer on the (0001) face.

Ribbons of any nominal orientation can in principle be produced. However, we have observed that straight steps may in some cases become rounded after annealing[17] and conversely etched pillars show faceting[3] along the armchair direction, as seen in Figure 6c-d. Rounding and faceting are sensitive to a number of factors such as step direction, growth condition (both temperature, time, and possibly other factors such as the type of heating) and the pinning of steps (for instance under amorphous carbon pads such as in figure 4b or by defects). The SiC trench height is a relevant parameter since shallow trenches (≤10nm) get washed out upon annealing and deep trenches tend to break into multiple facets revealing multiple parallel ribbons. Uniform and well formed graphene ribbons of various shapes are consistently realized on trenches in the range 15-35 nm deep.

An important point of consideration is the ribbon integrity for transport measurements. Growth of secondary ribbons may occur if there are transverse substrate steps that cross the trenches. Graphene growth on these substrate steps produces "side arms" on the main ribbon that have been found to affect the transport[38]. An example is given on the AFM topographic and friction images of Figure 7a-d. The substep features are clearly visible on the terraces of Figure 7a (bright color) and give rise to ribbons seen as dark lines (low friction) in Figure 7b. A single ribbon can be measured by selecting a segment with no side arms and etching away the surrounding, as was done in the low temperature measurement in Ref. [17]. An alternative to get very long pristine ribbon is to choose a step/trench orientation parallel to the local miscut. That way, the terraces are flat (see Figure 7d and AFM trace in Figure 7e and very straight ribbons with no side arms can be formed over hundreds of microns length as shown in Figure 7d-f.

Ribbons ~30-50nm wide show ballistic transport at room temperature [17]. This was unambiguously demonstrated in local four-point resistance measurements in UHV, showing a length independent resistance from 1 to 17 microns at room temperature. Importantly, ballistic transport was observed not only for specific graphene orientations but also for chiral and curved ribbons. In an early experiment [39], a conductive AFM tip was brought into contact with a curved ribbon 35nm wide and 1 µm long provided with a contact at both ends (Figure 8a). The tip is scanned over the sample and the tip to contact resistance is recorded. The resistance is minimum when the tip is in contact with the ribbons, as shown in the color plot of Figure 8b. The minimum resistance R on each scan line is plotted as a function of the distance L to the contact in Figure 8c. After subtracting the gold contact and line resistance, we find R=$R_{contact}$ +(ΔR/ΔL) L, with $R_{contact}$ =0.9 (h/ $e^2$) and ΔR/ΔL =0.56 (h/$e^2$)/µm. The contact resistance indicates there is only one conducting channel; in these conditions[40] the slope ΔR/ΔL=(h/ $e^2$)(L/λ) gives a mean free path λ=1.8µm. This value is longer that the ribbon length indicating ballistic transport. This result is remarkable because the ribbon is exposed to ambient conditions, which in general increases significantly the resistance in contrast to protected ribbons in UHV or encapsulated under a dielectric ($Al_2O_3$ like in Ref.17). We believe that scanning the tip over the ribbon cleanses the ribbon by sweeping away the contaminants. We have often observed accumulation of particles on the rougher buffer layer next to ribbons.

## 3. Large scale integration

By design, multiple ribbons can be produced all at once. Arrays of tens of thousands of ribbons have been grown on the sidewall of trenches patterned by photolithography[12]. An example of integration of 10,000 field effect transistors/$cm^2$ is presented in the optical image of Figure 9, where an aluminum gate (G) lies on a ribbon in between Pd/Au source and drain contacts (S - D).

Mainstream electronics and very large scale integration is based on silicon; it is therefore important to devise schemes to integrate graphene with Si wafers. The route we proposed[41] is based on the standard industry methods using wafer bonding and the smart-cut technique. The idea here is to transfer a Si film from a Si wafer onto a graphitized SiC wafer. Figure 9c shows the principle of the design. The top- bounded Si wafer is ready for device patterning, and the graphene layer beneath is accessed through metallic vias managed through the sub- micron thick Si wafer. This 3D integration realizes the interconnection of the SiC supported graphene platform and the Si-based electronic wafer while preserving the integrity of graphene. Details have been published elsewhere; the main steps consist in the fabrication of graphene structures (either from patterned epitaxial graphene or from template growth on SiC sidewalls), evaporation of an alumina film that serves as a bonding layer between the SiC and the oxidized Si wafer, wafer bonding, and finally splitting of the Si wafer (smartcut) that leaves a thin crystalline Si layer on top. The main advantage of the process is that the graphene growth temperature (above 1500°C) is not limited by the presence of silicon (melting point 1414°C), or the limited SiC quality like in SiC epilayers on Si; the Si wafer is fully accessible and the stacking of wafers increases the areal density inspired by the 3D stacked layers Very Large Scale Integration Technology.

Figure 9d-e shows the top Si film bonded on a SiC chip The purple color corresponds to the strongly bonded regions. The zoom out optical image of Figure 9e shows an array of ribbons grown on parallel sidewall trenches as seen from the SiC (transparent) side of the bonded wafers. Note that the ribbons structure is intact after bonding.

## Conclusion

We have reviewed the main characteristics of epitaxial graphene grown on SiC by the confinement controlled sublimation method and selective graphene growth techniques. Most recent results on structured growth demonstrate that graphene nanoribbons can be grown on the sidewall of either natural SiC steps or trenches etched in SiC. Ribbons of 20-40nm wide and more than 100μm long can be very well confined onto the sidewall and well isolated (no sidearms) from other graphene structures. Schemes for large scale integration of epitaxial graphene ribbons with Si wafer were also presented.

## Acknowledgements

Financial support is acknowledged from AFOSR and NSF under grants No FA9550-13-1-0217 and 1506006, respectively. Additional support is provided by the Partner University Fund from the French Embassy.

# Figure Captions

Figure 1. Growth of epigraphene by the confinement control sublimation method (adapted from Ref. [3])
(a) Principle of graphene growth by thermal decomposition of SiC at high temperature, Si escapes and is partially re-absorbed on the SiC surface. (b) CCS growth furnace with induction heating elements and graphite crucible with calibrated hole to contain the Si vapor. (c) Schematics of the graphene layer growth on the top 6H/4H-SiC(000-1) and bottom (000-1) surfaces.

Figure 2. Multilayer epigraphene on the C-face topography images.
(a) STM image: 4x4nm (From Ref. [23]). (b) STM image: 400x400nm. The blue square represents the size of the STM scan in (a) (From Ref. [23]). (c) AFM scan: 40µmx40µm. The white lines are graphene pleats. (d) 400nm long STEM image across areas of different moiré pattern, showing that the top layer is flat (rms<50pm) and continuous (From Ref. [24]) (e) Scanning Electron Microscopy image of a graphene pleat (From ref. [42]).

Figure 3. Monolayer graphene on the C-face.
(a) AFM images of large monolayer patches draping over steps. (b) Raman spectroscopy of the monolayer C-face graphene in (a); the black trace is the raw data, the red trace is the graphene spectrum once the SiC Raman peaks have been subtracted by the NFM method. (c)-(d) quantum Hall effect in monolayer graphene showing properties of the $\nu$=2 plateau: (c) low field onset (3Tesla) for a sample of $\mu$ = 39,800 cm$^2$/Vs and charge density n = 0.19×10$^{12}$cm$^{-2}$. (From Ref. [13])(d) high temperature quantization (n=0.87 ×10$^{12}$cm$^{-2}$, $\mu$ = 20,000 cm$^2$/Vs – From Ref. [32]). (e) High frequency transistor with a monolayer C-face graphene channel. Inset design of the dual gate transistor; main panel: the intercepts of $U^{1/2}$= 1 (slope 20 dB/dec) gives a maximum oscillation frequency ($f_{max}$ =70GHz) for a 100nm gate length. U is the Mason's unilateral gain (power gain). The current amplification $H_{21}$ versus frequency gives a cutoff frequency $f_T$ =110 GHz. (From Ref. [35])

Figure 4. Structuring SiC.
(a) AFM topographic image of 4H-SiC large terraces structured before growth, and step profile, showing straight steps and flat terraces 20 to 40µm wide. (b)SiC Step pinning under an evaporated amorphous carbon grid. After annealing at 1350°C, the steps accumulate at one side of the grid in each enclosure providing large terraces at a locations defined by the grid. (From Ref. [37])(c) AFM topographic profile within the grid (top) before and (bottom) after step bunching annealing, showing that a large terrace has developed.

Figure 5. Selected growth.
(a) Hall bar-shaped graphene multilayer (light grey) grown directly into shape using a AlN mask (dark) evaporated onto bare SiC. (Bottom) Raman map of the 2D graphene peak for the central area in (a), showing that graphene grows only in the unmask areas. From Ref. ([14])(b) Graphene growth under a Si-rich SiN mask (From Ref.[15]). -a- Optical image of the SiN pattern and -b- of subsequent MEG growth on SiC (graphene is light color) -c- Raman 2D map and -d- Raman 2D/G map, showing that graphene growth is promoted where the mask resided. Scale bar is 10 µm. (c) SEM images of 30nm thick GaN grown on 4H-SiC using punctured graphene as a mask (From Ref. [16]). No GaN grows on graphene, as exemplified on the right side of the image. Inset: SEM image of the epigraphene mask with 75 nm holes in it showing SiC (light dots). Scale bar in the inset is 200 nm.

Figure 6. Side wall growth
(a) Relative epigraphene to 4H/6H-SiC orientation and direction of trenches to grow ribbons of in principle defined zigzag or armchair direction. (b) cross-sectional TEM image of a 40nm high step covered with one layer graphene (From Ref. [43]). (c) 3D rendition of a facetted pillar after annealing (d) AFM topographic (top) and EFM (bottom) image of annealed pillars, showing that graphene is confined to the side walls (bright ring in the EFM images) (adapted From Ref. [3]).

Figure 7. Perfecting sidewall ribbons.
(a)-(b) Etched trenches with non flat bottom and top terraces. (a) AFM topographic and (b) Friction Force Microscopy (FFM) image (brown is low friction, yellow is the rippled, rougher buffer layer). Graphene grows on the main steps (vertical lines) as well as on the substeps on the terraces. (c)-(d) Straight isolated sidewall ribbons grown on bunched natural steps. (c) AFM and (d) FFM images for etched trenches with non flat bottom and top terraces. (e) Topographic profile along the red line in (c). (f) AFM 3D image of a long isolated ribbon grown on a step.

Figure 8. Electrical transport in sidewall ribbons.
(a) EFM image of an isolated epigraphene ribbon connected to two large graphene pads. (b) Tip to sample resistance color map showing that conduction is along the ribbon only (Resistance scale is on the right). (c) Plot of the resistance as a function of length along the ribbon in ambient condition.

Figure 9. Large scale integration
(a) Schematics of a field effect transistor (FET) with a sidewall ribbon as the channel connected to source S and drain D with a top gate G. (b) Integration of 10,000 FET/cm$^2$ (optical image) with parallel ribbons connected between Pd/Au contacts and provided with an aluminum gate (G) on an $Al_2O_3$ oxide (From Ref. [12]). (c) Principle of epigraphene on SiC to Si wafer bonding, with a top Si wafer ready for CMOS technology and an underneath epigraphene circuit. Both are interconnected by metal vias through the thin Si layer. (d-e) Realization of the epigraphene on SiC to Si wafer bonding. (d) optical image of the bi-wafer chip from the Si side. The purple color corresponds to the strongly bonded regions. (e) optical image from the transparent SiC side, showing an intact trench array supporting the sidewall ribbons. (From Ref. [41]).

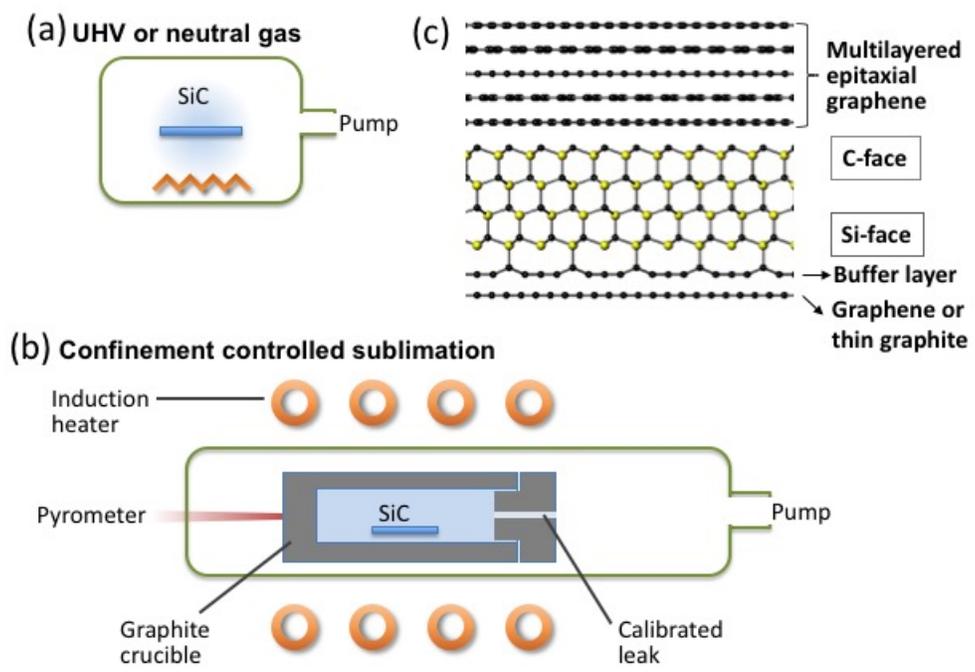

**FIGURE 1**

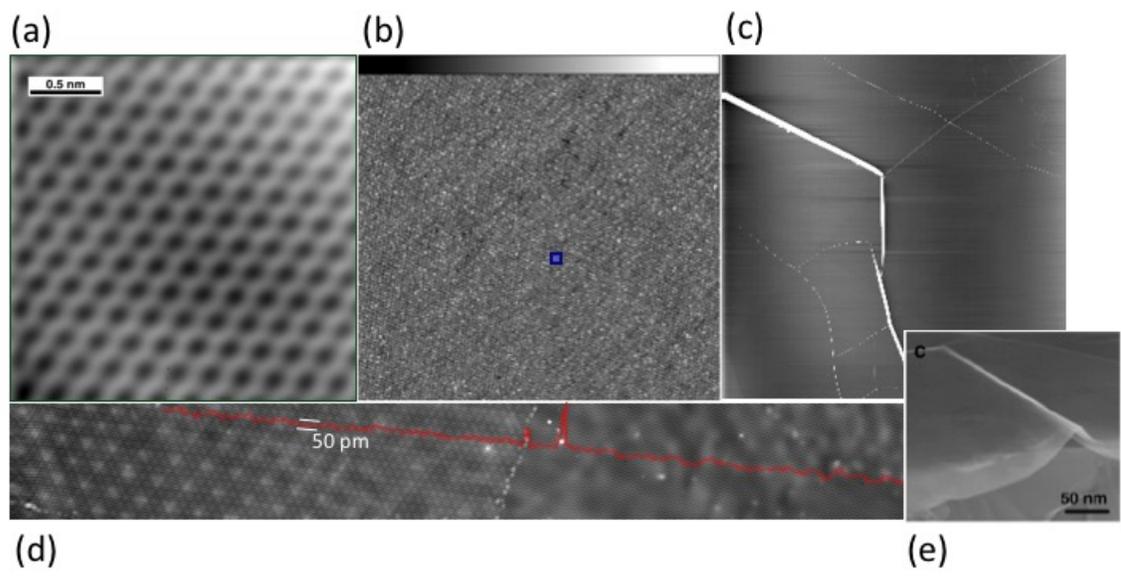

**FIGURE 2**

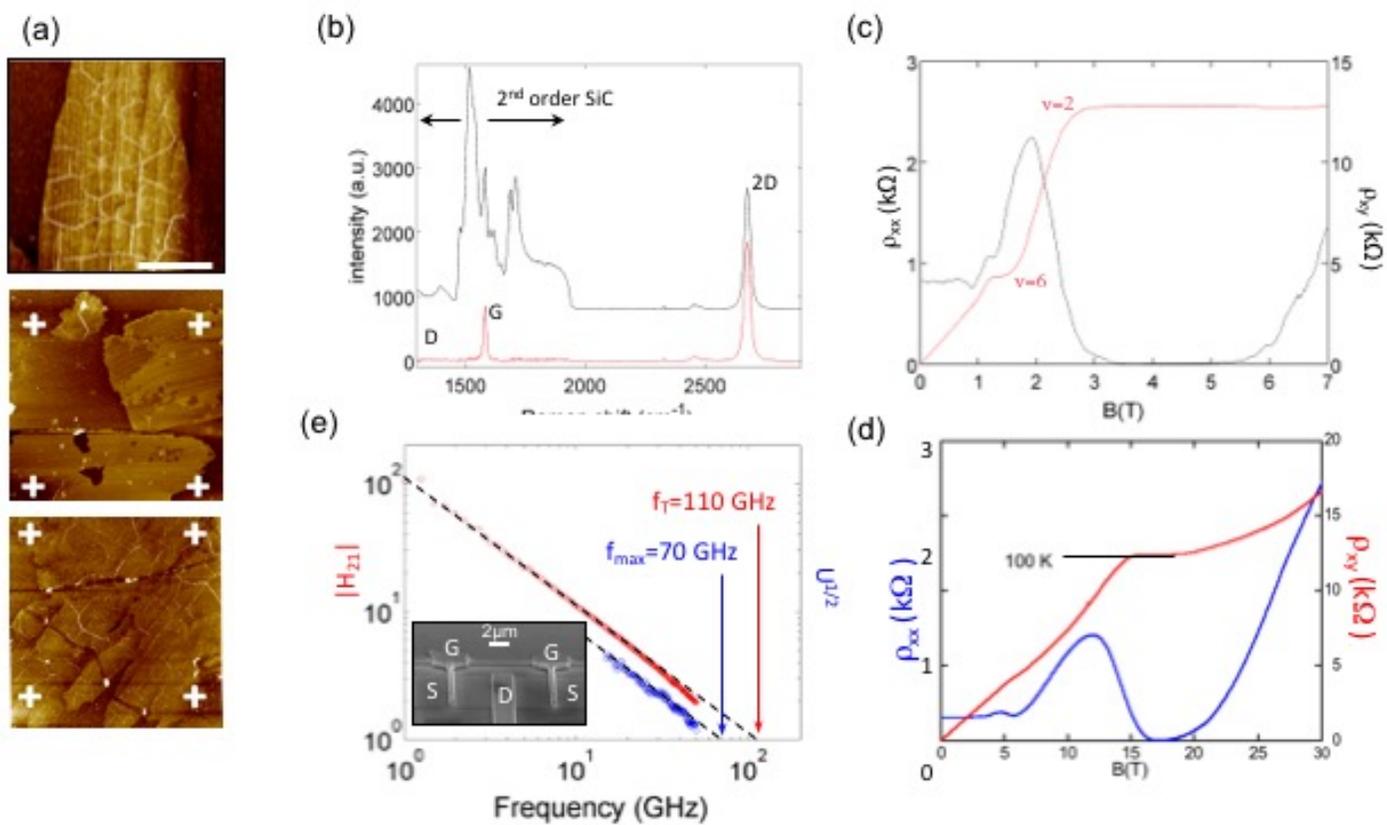

**FIGURE 3**

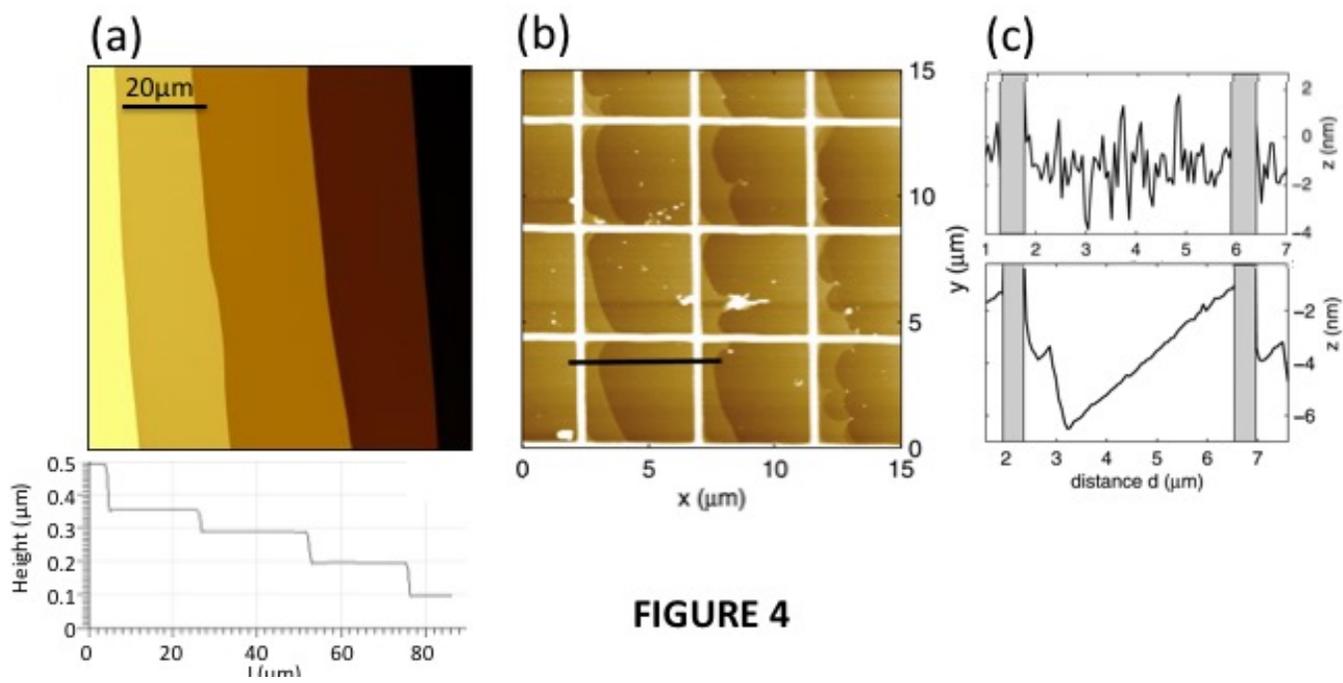

**FIGURE 4**

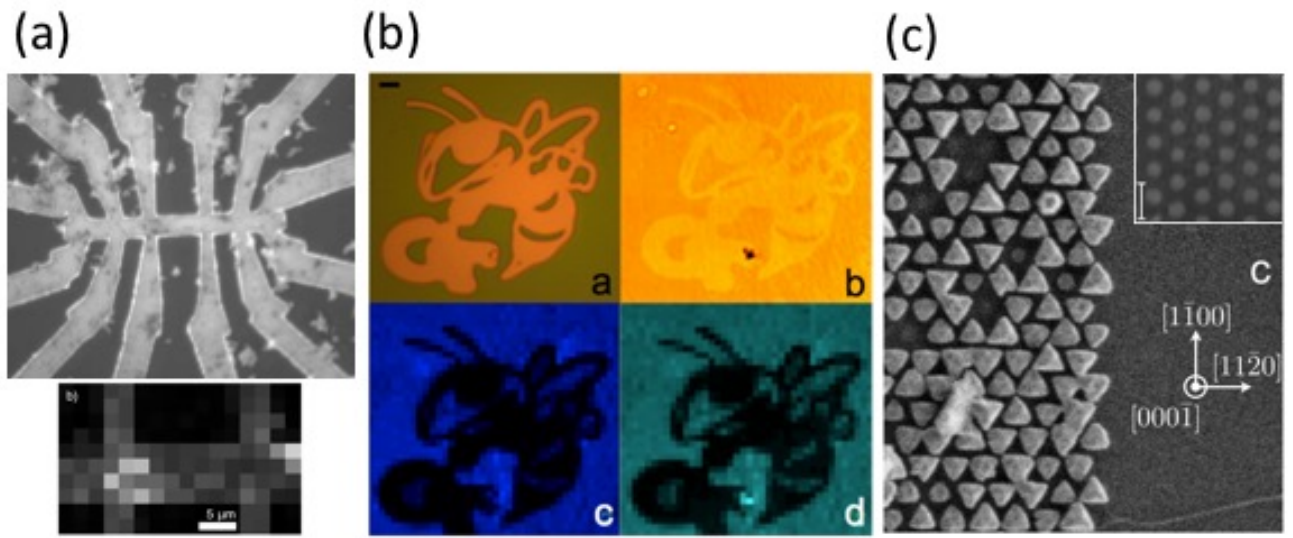

**FIGURE 5**

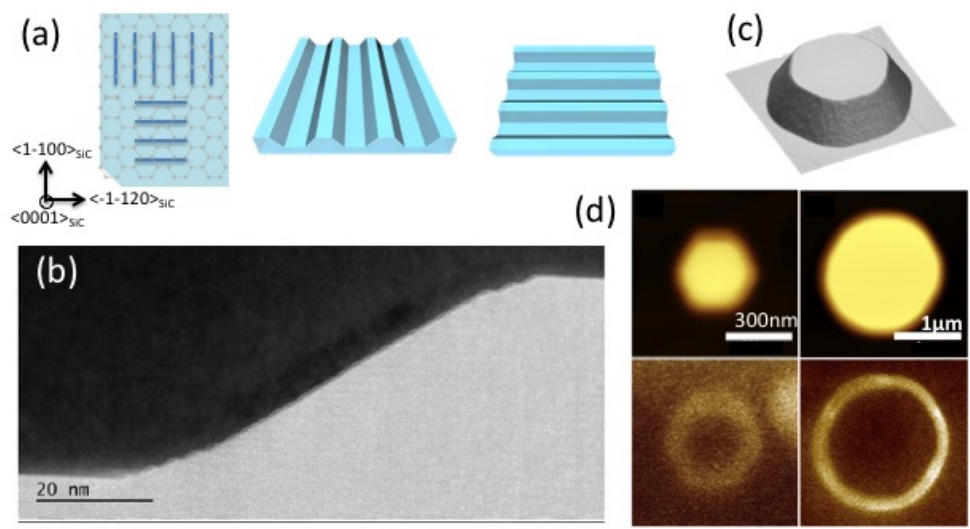

**FIGURE 6**

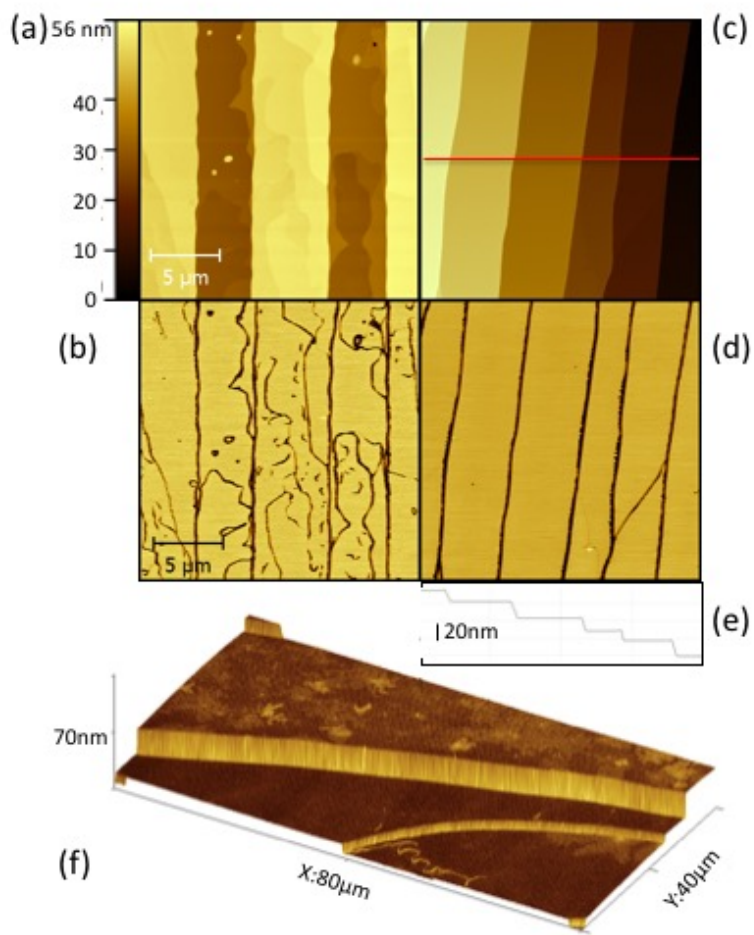

**FIGURE 7**

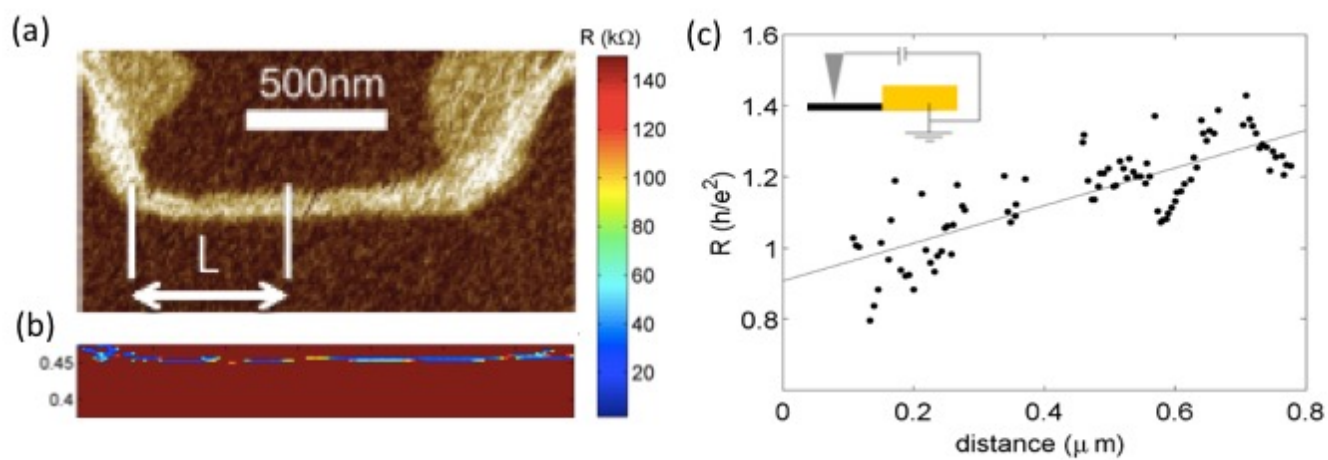

**FIGURE 8**

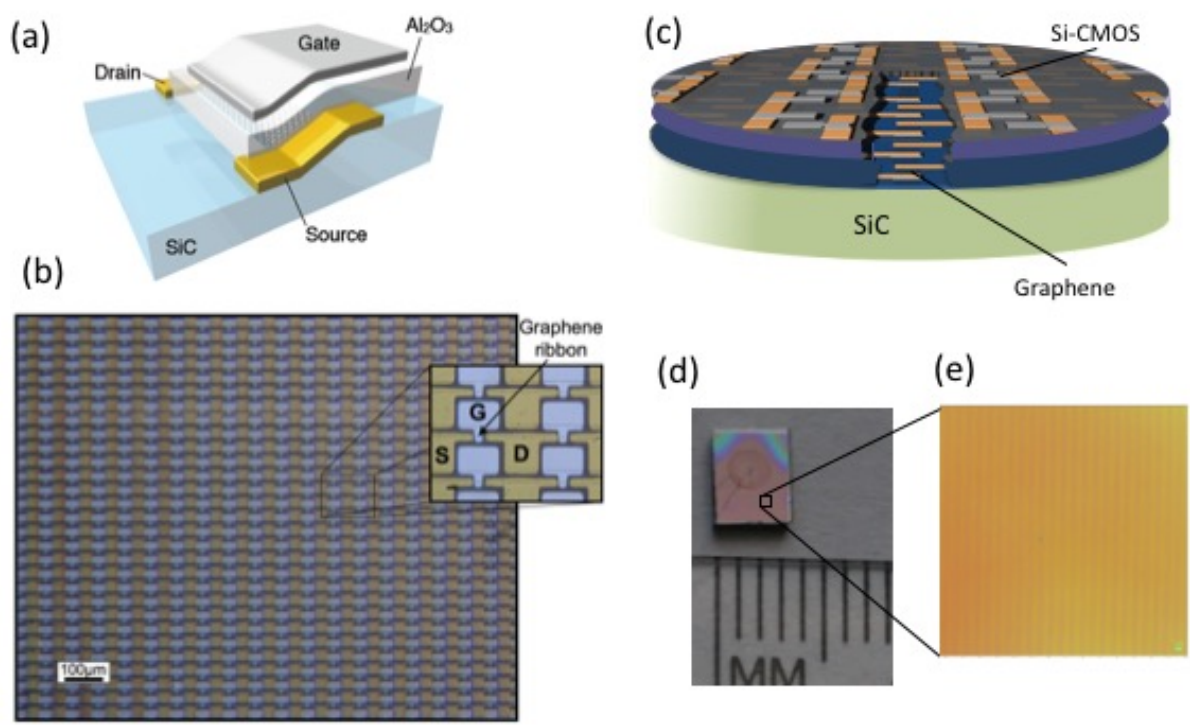

**FIGURE 9**